\documentclass[aps,prd,superscriptaddress,nofootinbib,preprintnumbers]{revtex4-2}

\usepackage{graphicx}
\usepackage{graphics}
\usepackage{amsmath,amssymb}
\usepackage[usenames]{color}
\usepackage{subfigure}
\usepackage{hyperref}
\usepackage{enumitem}
\usepackage{float}
\usepackage[titletoc]{appendix}

\def\Tr{\,{\rm Tr}\,}

\newcommand{\da}{\dagger}

\newcommand{\be}{\begin{equation}}
\newcommand{\ee}{\end{equation}}
\newcommand{\bea}{\begin{eqnarray}}
\newcommand{\eea}{\end{eqnarray}}
\newcommand{\ben}{\begin{enumerate}}
\newcommand{\een}{\end{enumerate}}
\newcommand{\bit}{\begin{itemize}}
\newcommand{\eit}{\end{itemize}}

\newcommand{\la}[1]{\label{#1}}

\newcommand{\Eq}[1]{Eq.~(\ref{#1})}

\newcommand{\Sec}[1]{Sec.~\ref{#1}}

\def\nl{\nonumber \\}

\newcommand{\vv}[1]{\mathbf #1}							% 3-vector
\newcommand{\bert}{\raise-0.45mm\hbox{\Large$\Box$}}			% D'Alembertian

\newcommand{\gd}{\gamma_\downarrow}						% Decay rate
\newcommand{\gu}{\gamma_\uparrow}						% Pumping rate

\makeatletter
\newcommand*\bigcdot{\mathpalette\bigcdot@{.5}}
\newcommand*\bigcdot@[2]{\mathbin{\vcenter{\hbox{\scalebox{#2}{$\m@th#1\bullet$}}}}}
\makeatother

\definecolor{BrickRed}{cmyk}{0,0.89,0.94,0.28}					%%%PANTONE 1805
\definecolor{MidnightBlue}{cmyk}{0.98,0.13,0,0.43}				%%%PANTONE 302
\definecolor{DarkGreen}{rgb}{0.100806,0.495968,0.209979}
\definecolor{orange}{rgb}{0.587167,0.354498,0.146197}

\graphicspath {{figures/}}

\begin{document}

\preprint{IFIC/23-31}

\title{Quantum thermodynamics of de Sitter space}

\author{Robert Alicki}
\email{robert.alicki@ug.edu.pl}
\affiliation{International Centre for Theory of Quantum Technologies (ICTQT), University of Gda\'nsk, 80-308, Gda\'nsk, Poland}
\author{Gabriela Barenboim}
\email{gabriela.barenboim@uv.es}
\affiliation{Departament de F\'isica Te\`orica and IFIC, Universitat de Val\`encia-CSIC, E-46100, Burjassot, Spain}
\author{Alejandro Jenkins}
\email{alejandro.jenkins@ug.edu.pl}
\affiliation{International Centre for Theory of Quantum Technologies (ICTQT), University of Gda\'nsk, 80-308, Gda\'nsk, Poland}
\affiliation{Laboratorio de F\'isica Te\'orica y Computacional, Escuela de F\'isica, Universidad de Costa Rica, 11501-2060, San Jos\'e, Costa Rica}

\date{First version: 10 Jul. 2023.  This revision: 12 Dec. 2023.  To appear in Phys.\ Rev. D}

\begin{abstract}
We consider the local physics of an open quantum system embedded in an expanding three-dimensional space $\vv x$, evolving in cosmological time $t$, weakly coupled to a massless quantum field.  We derive the corresponding Markovian master equation for the system's nonunitary evolution and show that, for a de Sitter space with Hubble parameter $h = $ const., the background fields act as a physical heat bath with temperature $T_{\rm dS} = h / 2 \pi$.  The energy density of this bath obeys the Stefan-Boltzmann law $\rho_{\rm dS} \propto h^4$.  We comment on how these results clarify the thermodynamics of de Sitter space and support previous arguments for its instability in the infrared.  The cosmological implications are considered in an accompanying letter \cite{relax}.  

 \end{abstract}

\maketitle

 \tableofcontents
 
 \newpage

%%%%%%%%%%
%%% INTRODUCTION
%%%%%%%%%%

\section{Introduction}
\la{sec:intro}

According to Einstein's theory of general relativity (GR), the size of our Universe is a dynamical quantity.  The space-time geometry of a homogenous, isotropic, and Euclidean universe can be described by the Friedmann-Lema\^itre-Robertson-Walker (FLRW) metric
\be
ds^2 = - dt^2 + a^2 (t) \, d \vv x^2 ,
\la{eq:FLRW}
\ee
where $(t, \vv x) = (t, x, y, z)$ are the space-time coordinates in the ``cosmic rest frame'' and $a(t) > 0$ is the scale factor that characterizes the size of the Universe.  In terms of the Hubble parameter
\be
h(t) \equiv \frac{\dot a (t)}{a (t)}
\la{eq:h}
\ee
(where the dot indicates derivation with respect to time $t$), Einstein's classical field equations for GR give us the two Friedmann equations:  First,
\be
h^2 = \frac{8 \pi \rho}{3} ,
\la{eq:1Fried}
\ee
where $\rho (t)$ is the energy density of the Universe as a function of cosmological time and, second,
\be
\dot h = -\frac 3 2 h^2 - 4\pi p ,
\la{eq:2Fried}
\ee
where $p$ is the pressure of the ``cosmological fluid'' with energy density $\rho$.  Equations \eqref{eq:1Fried} and \eqref{eq:2Fried} imply the condition of covariant energy-momentum conservation:
\be
\dot \rho = - 3 h (\rho + p) .
\la{eq:conserve}
\ee
In terms of the equation of state parameter $w \equiv p / \rho$, Eqs.\ \eqref{eq:1Fried} and \eqref{eq:conserve} imply that $h$ is constant if and only if $w = -1$.  Throughout this article we work in Planck units, such that \hbox{$G = M_{\rm Pl}^{-2} = \hbar = k_B = c = 1$}.

Hubble's astronomical observations in 1929 implied an expanding Universe ($\dot a > 0$) \cite{Hubble}.  This must have consequences for local physics.  Already in 1939, Schr\"odinger argued that an accelerated expansion ($\ddot a > 0$) should be accompanied by what he called ``alarming phenomena'' of particle production or annihilation \cite{Schroedinger}.  He drew this conclusion from the fact that such expansion mixes the positive and negative frequency terms in the Fourier expansion of a free relativistic field.  If the field is quantized, these terms are associated with production and annihilation of particles.  The question of ``cosmological particle production'' has remained contentious in theoretical physics because of the ambiguity about the choice of vacuum state and the associated definition of particle number (see, e.g., \cite{Ford-rev} and references therein).

In 1977, Gibbons and Hawking found that ``an observer moving on a timelike geodesic in de Sitter space [i.e., with $h=$ const.] will detect thermal radiation'' \cite{GH}.  They interpreted the temperature of this radiation, $T_{\rm dS} = h / 2 \pi$, as resulting from the presence of an event horizon in de Sitter (dS) space, similar to the horizon of the Schwarzschild solution that leads to Hawking radiation from a black hole.

The Hawking temperature for a black hole of mass $M$ is $T_H = 1 / (8 \pi M)$.  The total energy of a black hole can be identified with its mass $M$, so that the first law of thermodynamics takes the form
\be
dM = T_H dS + \Omega dJ + \Phi dQ ,
\la{eq:BH1}
\ee
where $S$ is the black-hole entropy, $J$ its angular momentum, $Q$ its charge, $\Omega$ its angular velocity, and $\Phi$ its electrostatic potential at the horizon.  This allows us to relate the Hawking temperature of the black hole to its entropy.  Assuming that $S = 0$ for $M=0$, \Eq{eq:BH1} implies that $S = 4 \pi M^2 = A_H / 4$, where $A_H$ is the area of black hole's event horizon.  This approach, however, does not work for dS space, which has no extensive variable to play the role of $M$ in \Eq{eq:BH1} \cite{LesHouches}.  If we take the dS entropy to be proportional to the surface area of the event horizon in the same way as the black-hole entropy, we get $S_{\rm dS} = \pi / h^2$, which diverges in the Minkowski limit ($h \to 0$) in which the horizon disappears.  How to interpret $T_{\rm dS}$ within a consistent thermodynamics for dS space remains contentious in high-energy theoretical physics: see, e.g., \cite{Anninos, deAlwis} and references therein.

In 1981, Guth proposed that a short-lived dS phase in the early Universe could solve the horizon and flatness problems in cosmology \cite{Guth}.  This idea, which Guth called ``inflation'', soon became a pillar of modern cosmology, especially after it was shown that it could also account for the primordial density perturbations that seeded the subsequent formation of galaxies; see, e.g., \cite{Weinberg} and references therein.  However, despite the acceptance of inflation as a general paradigm, none of the many detailed models that have been proposed to implement it have thus far been generally regarded as wholly satisfactory.  The main difficulties that any model of inflation faces are:

\begin{enumerate}[label=(\alph*)]

\item to account for the very special (i.e., low-entropy) initial state of the Universe before inflation,

\item to explain why inflation ended abruptly after the scale factor of the Universe had increased by $\gtrsim e^{60}$ (``60 e-folds''),

\item to explain how the energy that drove inflation was converted into thermal radiation, generating the observed entropy of our Universe, and 

\item to explain why accelerated expansion restarted long after the end of inflation, at a far slower rate of exponential growth \cite{Riess, Perlmutter}.

\end{enumerate}

In this article we study the thermodynamics of dS space using analytical techniques based on the Markovian master equation (MME) for an open quantum system (also known as the ``GKLS equation'' or the ``Lindblad equation'') \cite{GKS, Lindblad}.  Taking such a system to be weakly coupled to a background massless quantum field and working in the cosmic rest frame for dS space, we find that the system equilibrates to a population distribution with temperature $T_{\rm dS} = h / 2 \pi$, consistent with the result of \cite{GH}.  The authors of \cite{GH} concluded, based on the dS isometries, that any other observer moving along a timelike geodesic would also measure $T_{\rm dS}$.  We find, however, that this temperature is only well defined in the cosmic rest frame, in which the background acts as a {\it physical} heat bath whose energy density obeys the Stefan-Boltzmann law $\rho_{\rm dS} \propto T_{\rm dS}^4$.\footnote{We have borrowed the term ``cosmic rest frame'' from the astrophysical literature, where it is commonly used to label the preferred frame in which the cosmic microwave background (CMB) has no dipole moment.  In that case, it is well understood that the CMB has a blackbody spectrum ---and therefore a well defined temperature--- only in the cosmic rest frame.  See, e.g., \cite{LL-CRF}.}

In an accompanying letter \cite{relax}, we combine this result with the Friedmann equations of classical GR and find that the dS phase ends abruptly because of the irreversible transfer of energy from the background $\rho_{\rm dS}$ into ordinary particles.  This provides a graceful exit to inflation without the need to invoke any inflaton potential.  Thermal particle production during inflation, with a blackbody spectrum given by $T_{\rm dS}$ (rather than vacuum fluctuations) can explain the presence of adiabatic, Gaussian, and approximately scale-invariant primordial perturbations that seed the subsequent formation of structure in the Universe.

Various authors have already argued that dS space is unstable in the infrared (IR) and that such an instability may explain how inflation ends and why the cosmological constant is currently so small compared to the natural scale $M_{\rm Pl}^4$; see, e.g., \cite{Myhrvold, Ford-IR, Mottola, Antoniadis, Tsamis-Woodard96, Tsamis-Woodard98, Polyakov07, Polyakov12, Dvali-Gomez, Akhmedov} (this list is far from exhaustive).  However, our limited understanding of the thermodynamics of dS space has obscured the logical relations among these various arguments and prevented any of them from being widely accepted as decisive.  This work seeks to apply to that problem the analytical methods of open quantum systems and quantum thermodynamics.  Such an approach has, thus far, been used mostly in quantum optics (for an introduction to quantum thermodynamics and its history, see \cite{QT}).  The resulting formulation of the local dynamics in dS space is not generally covariant.  This is a feature and not a bug of our approach.  It is well established ---though perhaps not very widely appreciated--- that thermodynamic quantities are necessarily {\it not} covariant in quantum physics \cite{Sewell} (see also the discussion in the Appendix).  Moreover, it has been shown explicitly that a finite entropy breaks the classical space-time symmetries of dS space \cite{trouble}.

The combined results of this article and of \cite{relax} have direct relevance to points (b) and (c) in the above list of difficulties faced by theories of inflation.  Our analysis may also have something to contribute to point (a), as we comment upon briefly in \Sec{sec:IR}, but this point remains speculative.  We have nothing to contribute on point (d) (the ``small cosmological constant problem''), except to the extent that our results could be combined with further efforts of model building aimed specifically at accounting for the current phase of slowly accelerating expansion.

%%%%%%%%%%
%%% RELATION TO OTHER APPROACHES
%%%%%%%%%%

\section{Relation to other approaches}
\la{sec:other}

The theoretical literature on dS thermodynamics and related problems in particle physics and cosmology is vast.  Some further remarks are therefore in order about how our work relates to what other researchers have done.  Starobinsky pioneered the study of the stochastic dynamics of the average of a scalar field over a finite region of dS space \cite{Starobinsky}.  More recently, Mirbabayi has studied the dynamics of such semiclassical averages over the causal wedge of a dS observer and found that they equilibrate to a thermal state in a Markovian fashion \cite{Mirbabayi}.  Our own approach, based on the MME formalism, is very different because we consider the {\it local} dynamics of an open quantum system embedded in dS space and show that, if the system is weakly coupled to a massless field, then an observer in the cosmic rest frame sees this local system thermalize to the Gibbons-Hawking temperature.  Thus, no spatial averaging is involved in our formulation.  Understanding the relation between the results of these two qualitatively different approaches is an interesting problem that we must leave for future investigation.

Chandrasekaran, Longo, Penington, and Witten have recently considered the algebra of observables for a static patch in dS space, thereby emphasizing that physical measurements carried by an observer in dS are affected by the expansion of the space in which the observer sits \cite{CLPW}.  Susskind has made a qualitatively similar point by invoking the concept of ``quantum reference frames'' \cite{Susskind}.  Although our formulation of dS thermodynamics based on the MME is quite different from those of \cite{CLPW} or \cite{Susskind}, it does share with them a focus on understanding how the dS background affects the physics of the local quantum systems with which observers probe their surroundings.  In our case, the probe will act as a thermometer, as we will discuss in detail in \Sec{sec:dS-QFT}.

The MME has been applied previously to early Universe cosmology: see, e.g., \cite{Burgess-etal, Boyanovsky, Hollowood-McDonald, Martin-Vennin, Brahma-etal, Colas-etal} and references therein.  As far as we know, however, it has only been used to describe decoherence effects and not the thermal processes that we will be interested in here.  We hope that future research will connect the results of this paper to ongoing work by others on how to use the MME formalism to understand irreversible process in the early Universe.

%%%%%%%%%%
%%% LOCAL PHYSICS IN EXPANDING SPACE
%%%%%%%%%%

\section{Local physics in expanding space}
\la{sec:local}
 
We seek to describe the irreversible dynamics of an open quantum system embedded in an expanding three-dimensional space $\vv x = (x, y, z)$ and evolving in cosmological time $t$.  If the system in static space is characterized by the Hamiltonian $\hat H$, then its dynamics in an expanding space can be expressed by the modified Hamiltonian
\be
\hat H_D (t) = \hat H +  h(t) \hat D ,
\la{eq:H}
\ee
where $\hat D$ is the spatial dilation operator and $h(t)$ is the Hubble parameter defined in \Eq{eq:h}.  This separation of $\hat H$ is valid if we may neglect the backreaction of the dynamics of the system on the space-time, and it is therefore equivalent to the fundamental approximation made in calculations of quantum fields in curved space-time (see, e.g., \cite{Jacobson, Parker-Toms} and references therein).  In the accompanying letter \cite{relax} we include the backreaction via an adiabatic approximation, by taking $h(t)$ from the classical Friedmann Eqs.\ \eqref{eq:1Fried} and \eqref{eq:2Fried}, which is consistent with the usual approach in cosmology.

This formulation is not coordinate invariant, but it will allow us to apply the analytical machinery of quantum thermodynamics in a straightforward way.  Moreover, in cosmology we are primarily interested in how the dynamics appears to an observer in the cosmic rest frame of \Eq{eq:FLRW}.  The formulation of \Eq{eq:H} is analogous to that of previous work in which the quantum thermodynamics of moving baths (including black-hole superradiance) was studied using an effective Hamiltonian shifted by rotation \cite{rotating}, with the important difference that in that case $\hat H$ was taken to commute with the generator of rotations $\hat L_z$, whereas now we do not take $\hat H$ to commute with $\hat D$.

For a wave function $\psi(\mathbf{x})$ the operator $\hat{D}$ is defined by
\be
\left[ e^{-i \lambda \hat D} \psi \right] (\mathbf{x} ) = e^{- \frac 3 2 \lambda} \psi \left( e^{-\lambda} \mathbf{x} \right) ,
\la{eq:D}
\ee
and hence
\be
\hat D =  \frac 1 2 \left( \hat{\vv x} \cdot \hat{\vv p} + \hat{\vv p} \cdot \hat{\vv x} \right) ,
\la{eq:Dgen}
\ee
which is a particular case of the ``squeeze operator''.  The Hamiltonian of \Eq{eq:H} for a single nonrelativistic particle is
\be
\hat H_D (\hat{\vv x}, \hat{\vv p}; t) =  \frac{\hat{\vv p}^2}{2m}+ U(\hat{\vv x}) + \frac 1 2 h(t) \left(\hat{\vv x} \cdot \hat{\vv p} +\hat{\vv x}\cdot \hat{\vv p} \right) .
\la{eq:H1bod}
\ee
Under the unitary transformations $\hat{\vv x} \mapsto \hat{\vv x}$ and $\hat{\vv p} \mapsto \hat{\vv p} + h m \hat{\vv x}$ this becomes
\be
\hat H'_D (\hat{\vv x}, \hat{\vv p}; t) =  \frac{\hat{\vv p}^2 }{2m}+ U(\hat{\vv x}) - \frac 1 2 m h^2(t) \hat{\vv x}^2 
\la{eq:H1bod-eq}
\ee

As a check that \Eq{eq:H} gives the correct local physics in an expanding Universe, consider a two-body system governed by the classical version of Eqs.\ \eqref{eq:H} and \eqref{eq:Dgen}:
\be
H(\vv x, \vv p; t) =  \frac{\vv p_A^2 }{2m_A}+ \frac{\vv p_B^2 }{2m_B} + U(\vv x_A -\vv x_B) + h(t) \left(\vv x_A \cdot \vv p_A +
 \vv x_B \cdot \vv p_B \right) .
\la{eq:H2bod}
\ee
Using the index $K= A,B$, the corresponding Hamiltonian equations can be written as
\be
\dot{\vv x}_K = \frac{\partial H}{\partial \vv p_K} =\frac{\vv p_K}{m_K} + h(t) \vv x_K , \quad \dot{\vv p}_K =-\frac{\partial H}{\partial \vv x_K} = -\frac{\partial}{\partial \vv x_K} U(\vv x_A - \vv x_B) - h(t) \vv p_K .
\la{eq:Heq}
\ee
From \Eq{eq:Heq} one obtains the Newtonian equations of motion
\be
m_K \ddot{\vv x}_K =  -\frac{\partial}{\partial \vv x_K} U(\vv x_A - \vv x_B) + m_K \left[ h^2(t) + \dot h(t) \right] \vv x_K .
\la{eq:Newton}
\ee
Introducing the center of mass coordinate $\vv X = M^{-1}(m_A \vv x_A + m_B \vv x_B)$, $M= m_A + m_B$, and the relative position  $\vv x = \vv x_A -  \vv x_B$ this becomes
\begin{align}
\ddot{\vv X} &= \left[ h^2(t) + \dot h(t) \right] \vv X , \la{eq:NewtonX} \\
\ddot{\vv x} &= -\frac 1 \mu \frac{\partial}{\partial \vv x} U(\vv x) + \left[ h^2(t) + \dot h (t) \right] \vv x, \quad \mu= \frac{m_A m_B}{m_A + m_B } .
\la{eq:Newtonx}
\end{align}
For $h =$ const. (dS space) the contribution from expansion to \Eq{eq:Newtonx} takes the form of a negative (unstable) harmonic potential.  This agrees with the  ``all or nothing'' picture presented in \cite{Price-Romano, Faraoni-Jacques}: a tightly bound classical system will, after some initial disturbance of its orbit, evolve with bounded $| \vv x |$, unaffected by the continuing expansion of space, while the center of mass follows the Universe's expansion.

Note that there is an obvious canonical quantization of the system described by \Eq{eq:Newtonx}.  For a tightly bound system we expect that bound states are only slightly perturbed by the dilation. However, none of them will be eigenvectors of the total Hamiltonian with the dilation term.  Instead, they will be resonances that decay by tunneling through the potential barrier.  The lifetime of such a resonance can be very long if the system is tightly bound compared to the rate of expansion, but this already suggests that quantum systems in dS space are unstable in the infrared.  This is a crucial issue that we shall return to in \Sec{sec:IR} in light of the analytical results of quantum thermodynamics.

%%%%%%%%%%
%%% MARKOVIAN MASTER EQUATION
%%%%%%%%%%

\section{Markovian master equation}
\la{sec:MME}

Since this is a new application of an approach to thermal physics that has not been widely used in high-energy physics or cosmology (see \Sec{sec:other}), in this section we define the main mathematical objects used in the rest of this article, based on a concrete example: a simple harmonic oscillator weakly coupled to a thermal bath of bosons.  The general analytical methods and results from the theory of open quantum systems relevant to the present discussion are briefly reviewed in the Appendix.  For further details, see \cite{Alicki-Lendi} and references therein.

%%%%%%%%%%
%%% EXAMPLE: HARMONIC OSCILLATOR IN A BOSONIC BATH
%%%%%%%%%%

\subsection{Example: Harmonic oscillator in a bosonic bath}
\la{sec:bath}

Let us begin by reviewing the results of the MME applied to a harmonic oscillator weakly coupled to a bosonic bath.  The purpose of this exercise is to introduce the relevant concepts and analytical techniques in an intuitive and physically well motivated context.  Those same concepts and techniques will then be applied in \Sec{sec:dS-QFT} to the thermodynamics of dS space.

Take the physical Hamiltonian of a simple harmonic oscillator,
\be
\hat H = \omega \hat b^\da \hat b  ,\quad \mathrm{with} \quad \left[ \hat b , \hat b^\da \right] = 1 ,
\la{eq:Hosc}
\ee
where $\omega$ is the renormalized frequency of the oscillator (we ignore the energy of the ground state, which will not be relevant to us here).  See the Appendix for a discussion of the relation between the renormalized $\hat H$ and the bare Hamiltonian $\hat H_0$.  This oscillator serves as the simplest model of a localized system that can probe its surroundings.  In particular, we will consider how it evolves if it is weakly coupled to a much larger system that can be treated as a thermal bath.

Let us take the thermal bath to consist of noninteracting bosons with dispersion relation $\omega(\vv k)\geq 0$ and occupying an infinite three-dimensional space.  The bath Hamiltonian,
\be
\hat H_B = \int d^3 k\, \omega(\vv k) \hat a^\da (\vv k) \hat a (\vv k) ,
\la{eq:HB}
\ee
is given in terms of annihilation and creations field operators satisfying the canonical commutation relation:
\be
\left[ \hat a (\vv k) , \hat a^\da (\vv k') \right] = \delta^{(3)}(\vv k -\vv k') .
\label{eq:CCR}
\ee
The density of states (per unit volume) at a given energy $\omega$ is
\be
n(\omega) = \int \frac{d^3 k}{(2 \pi)^3} \delta [\omega - \omega(\vv k)] .
\la{eq:nd}
\ee
The equilibrium state of a bosonic gas at the inverse temperature $\beta$ is fully characterized by the second-order correlations
\be
\left \langle \hat a^\da (\vv k) \hat a (\vv k') \right \rangle_\beta  = \frac{1}{e^{\beta \omega (\vv k)} - 1} \delta^{(3)}(\vv k -\vv k') ,
\la{eq:BE}
\ee
where $\langle \cdots \rangle_\beta$ denotes the average with respect to equilibrium state with inverse temperature $\beta$.  Note that here we have simply {\it assumed} that the bosons are in the thermal state characterized by \Eq{eq:BE}.

The coupling of the harmonic oscillator to the bath is given by the interaction Hamiltonian
\be
\hat H_{\rm int} = \lambda \left( \hat b + \hat b^\da \right) \otimes \hat \phi_\Lambda (0) ,
\la{eq:Hint}
\ee
Here, the harmonic oscillator is localized at the origin of the coordinate system and locally coupled to the field at that point. To obtain a mathematically well-defined Hamiltonian we regularize the quantum field at the point using an ultraviolet cutoff parameter $\Lambda$, such that
\be
\hat \phi_\Lambda (0) = \int \frac{d^3 k}{\sqrt{2\omega(\vv k)}} e^{-\omega(\vv k)/\Lambda} \left[ \hat a (\vv k) + \hat a^\da (\vv k) \right] .
\la{eq:regfield}
\ee
Under the standard assumptions of weak coupling and of large separation between the slow time scale of the system and the fast time scale of the relaxation of the internal correlations in the bath, one obtains the MME for the nonunitary evolution of the density matrix $\hat \rho(t)$ corresponding to the mixed state of the system, with the degrees of freedom of the bath averaged over their equilibrium state.\footnote{In the rest of this article we will simply assume that the coupling between the localized system (here, the harmonic oscillator) and the bath (here, the bosonic field) is such that the Markovian approximation is valid.  Otherwise, the separation of the full physics into a small system of interest and a large environment acting as a thermal bath would be inappropriate.  In other words, we would not have a valid thermometer and, therefore, a thermodynamic treatment would not be useful.}  In terms of the renormalized Hamiltonian $\hat H$, this MME is
\be
\frac{d}{dt}\hat\rho (t)= -i \left[ \hat H , \hat\rho (t) \right] +\frac 1 2 \left( \gd \left( \left[ \hat b, \hat \rho(t) \hat b^\da \right] + \left[ \hat b \hat \rho(t), \hat b^\da \right] \right) + \gu \left( \left[ \hat b^\da, \hat \rho(t) \hat b \right] + \left[ \hat b^\da \hat \rho(t), \hat b \right] \right) \right)                                  
\la{eq:MME}
\ee
with damping and pumping rates
\be
\gd = \lambda^2\tilde G (\omega) \quad \hbox{and} \quad \gu = \lambda^2 \tilde G (-\omega)
\la{eq:rates}
\ee
expressed in terms of the {\it spectral density} $\tilde G (\omega)$ for the renormalized frequencies $\omega$.  This spectral density is given by the Fourier transform of the {\it bath correlation function} $G(t)$ (see Appendix).

%%%%%%%%%%
%%% SPECTRAL DENSITY FUNCTION
%%%%%%%%%%

\subsection{Spectral density function}
\la{sec:spectral}

In this case we have
\begin{align}
\tilde G (\omega) &= \lim_{\Lambda\to\infty}\tilde G_\Lambda (\omega) = \lim_{\Lambda\to\infty}\int_{-\infty}^{+\infty}dt\, e^{i\omega t} G_{\Lambda}(t) =  \lim_{\Lambda\to\infty}\int_{-\infty}^{+\infty}dt\, e^{i\omega t} \left\langle \hat \phi_\Lambda (t) \hat \phi_\Lambda (0) \right\rangle_{\beta}  \nl
&= \lim_{\Lambda\to\infty}\int_{-\infty}^{+\infty}dt\, e^{i\omega t} \left\langle e^{i\hat H_B t}\hat{\phi}_{\Lambda}(0)e^{-i\hat H_B t}\hat{\phi}_{\Lambda} (0) \right\rangle_{\beta}  \nl
&=\lim_{\Lambda\to\infty}  \int d^3 k\,  \int_{-\infty}^{+\infty}dt\, e^{i\omega t}\frac{e^{-2\omega(\vv k)/\Lambda}}{2\omega(\vv k)} \left\{\left\langle \hat{a}(\vv k)\hat{a}^{\dagger}(\vv k)\right\rangle_\beta e^{-i\omega(\vv k)t} +\left\langle \hat{a}^{\dagger}(\vv k)\hat{a}(\vv k) \right\rangle_\beta e^{i\omega(\vv k)t}  \right\}
 \nonumber \\
&=\pi  \int d^3 k\, \frac{1}{\omega}\left\{\delta[\omega - \omega(\vv k) ]\frac{1 }{1- e^{-\beta\omega}}+\delta[\omega + \omega(\vv k)] \frac{1 }{e^{\beta\omega }-1}\right\}  .
\la{eq:spectral}
\end{align}
Using \Eq{eq:nd} one obtains
\be
\tilde G (\omega)= \frac{8 \pi^4 n(|\omega|)}{\omega (1- e^{-\beta\omega })}.
\la{eq:spectralfin}
\ee
One can check the Kubo-Martin-Schwinger (KMS) property of the spectral density (see Appendix)
\be
\tilde{G}(-\omega)= e^{-\beta\omega} \tilde{G}(\omega) ,
\label{eq:KMS}
\end{equation}
which implies that the Gibbs state $\hat\rho_{\beta}= Z^{-1} e^{-\beta\hat H}$ is an equilibrium state for the harmonic oscillator and, moreover, that any initial state relaxes to $\hat\rho_{\beta}$.  The harmonic oscillator can therefore be treated as a thermometer that measures the temperature of the bosonic bath.

%%%%%%%%%%
%%% NOISE SPECTRA
%%%%%%%%%%

\subsection{Noise spectra}
\la{sec:noise}

The bosonic heat bath at equilibrium is a source of thermal/quantum noise characterized by two different spectra associated with $\tilde{G}(\omega)$.  The first one is the \emph{field fluctuation spectrum} $\mathcal{P}(\omega)$ defined as
\be
\int_0^{\infty} d \omega \, \mathcal{P}(\omega) = \lim_{\Lambda \to \infty} \langle \hat{\phi}^2_{\Lambda} \rangle_{\beta} .
\la{eq:Pdensity}
\ee
This average can be written as
\begin{align}
\langle \hat{\phi}^2_{\Lambda} \rangle_\beta &=
\int_0^\infty d \omega \int d^3 k\,  \delta[\omega-\omega(\vv k) ]e^{-2\omega(\vv k)/\Lambda}\frac{\left\langle \hat{a}^{\dagger}(\vv k)\hat{a}(\vv k) +  \hat a (\vv k) \hat a^\da (\vv k) \right\rangle_{\beta}}{2\omega(\vv k)} \nl
&= (2\pi)^3 \int_0^{\infty} d\omega \, \frac{e^{-2\omega/\Lambda}}{\omega} n(\omega) \left( \frac{1}{e^{\beta\omega}-1} + \frac 1 2 \right) .
\la{eq:Pdensity1}
\end{align}
This can be decomposed into thermal and vacuum contributions:
\be
\mathcal{P} (\omega) =  \mathcal{P}_{\rm th}(\omega)+ \mathcal{P}_{\rm vac}(\omega) , \quad \mathcal{P}_{\rm th}(\omega)= (2\pi)^3\frac{n(\omega)}{\omega(e^{\beta\omega}-1)}, \quad  \mathcal{P}_{\rm vac}(\omega)= 4\pi^3\frac{n(\omega)}{\omega} .
\la{eq:Pdensity2}
\ee
The $\mathcal{P}_{\rm vac}$ can be renormalized to zero by using normal ordering for the powers of field, as is usually done in statistical mechanics. 

The second spectrum associated with $\tilde{G}(\omega)$ is the \emph{energy density spectrum} $\mathcal{U}(\omega)$, given by 
\be
\mathcal{U} (\omega) = \frac {n(\omega)\, \omega}{e^{\beta\omega} -1} .
\la{eq:energydensity}
\ee
This coincides with the Planck distribution (per single polarization) for massless scalar bosons with $\omega(\vv k) = |\vv k|$ and density of states
\be
n(\omega) = \frac{\omega^2}{2 \pi^2} .
\la{eq:nPl}
\ee
The two spectra are related by
\be
\mathcal{P}_{\rm th}(\omega)= 2\pi \tilde G (-\omega) \quad \hbox{and} \quad \mathcal{U}(\omega) = \frac{\omega^2}{(2\pi)^3} \mathcal{P}_{\rm th}(\omega) = \frac{\omega^2}{(2\pi)^2}\tilde G (-\omega) .
\la{eq:Pdensity3}
\ee

%%%%%%%%%%
%%% THERMALIZATION IN EXPANDING UNIVERSE
%%%%%%%%%%

\section{Thermalization in expanding Universe}
\la{sec:dS-QFT}

It should be clear from the discussion in \Sec{sec:MME} that our main task, in order to describe the thermodynamics of dS space analytically, is to find the corresponding spectral density $\tilde G (\omega)$, which is given by the Fourier transform of the bath correlation function $G(t)$ [see \Eq{eq:spectral}].  To obtain $G(t)$ we need to know how the operator $\hat \phi_\Lambda (0)$ of \Eq{eq:regfield} evolves in cosmological time $t$, for an expanding space $\vv x$ characterized by the Hubble parameter $h(t)$.  It is important to emphasize that, unlike in \Sec{sec:MME}, here we will not assume any thermal state.  Instead, we will show that in dS space the quantum vacuum acts as a thermal bath, and we will compute its temperature and noise spectra.  Our treatment is similar in spirit to the derivation of the Unruh effect presented in \cite{Feng-Zhang}.

\subsection{Time-evolution operator}
\la{sec:propagator}

Let us consider a scalar field described in the cosmic rest frame at some initial cosmological time (say, $t_0 = 0$) by annihilation and creation operators $\hat a(\vv k) ,\hat a^\da(\vv k)$ satisfying the commutation relations of \Eq{eq:CCR}. The reference state of the field is the vacuum $| \Omega \rangle$ satisfying $\hat a(\vv k) | \Omega \rangle = 0$, which is consistent with the Bunch-Davies condition for the dS vacuum \cite{BD}.  The effects of the expansion of space are accounted for by adding a time-dependent dilation generator to the free Hamiltonian, as in \Eq{eq:H}.  For a single massless scalar boson we have
\be
\hat H_D(t) = \hat H + h(t) \hat D , \quad \hat H = |\vv k|, \quad  \hat D = i \left( \vv k \frac{\partial}{\partial \vv k} + i \frac 3 2 \right) .
\la{eq:H1a}
\ee
In order to solve the corresponding Schr\"odinger evolution equation we use the commutation relation
\be
\left[ \hat D, \hat H \right] = i \hat H .
\la{eq:commHD}
\ee
By the Baker-Hausdorff lemma, this implies that
\be
e^{-i\alpha \hat D} \hat H  e^{i \alpha \hat D} = e^\alpha \hat H .
\la{eq:HD-BH}
\ee
The time-evolution operator for a single boson can be written as
\be
\hat U(t) = \mathbf{T} \, \exp{\left[ -i \int_0^t ds \, \left( \hat H + h(s) \hat D \right) \right]} = e^{-i \nu(t) \hat D} e^{-i \tau(t) \hat H} .
\la{eq:U-wave}
\ee
The time-ordering symbol $\mathbf T$ is needed if $h(t)$ is not constant, since then
\be
[\hat H_D (t_1), \hat H_D (t_2)] = i [ h(t_1) - h(t_2) ] \hat H \neq 0 \quad \hbox{for} \quad t_1 \neq t_2 .
\la{eq:noncomm}
\ee

To show that $\hat U(t)$ does indeed take the form given in \Eq{eq:U-wave}, we compute its derivative in time:
\be
\hat U'(t) = - i \nu'(t) \hat D \hat U(t) + e^{-i \nu (t) \hat D} \left( - i \tau'(t) \hat H \right) e^{-i \tau (t) \hat H}
= -i \left[ \nu'(t) \hat D + \tau'(t) e^{-i \nu(t) \hat D} \hat H e^{i \nu(t) \hat D} \right] \hat U(t) .
\la{eq:dU}
\ee
Using \Eq{eq:HD-BH}, we conclude that, as long as
\be
\nu(t) =  \int_0^t ds \, h(s) \quad \hbox{and} \quad \tau(t) = \int_0^t ds \, e^{-\nu (s)} ,
\la{eq:newtime}
\ee
this $\hat U(t)$ obeys the Schr\"odinger equation
\be
\hat U'(t) = -i \hat H_D (t) \, \hat U(t) \quad \hbox{for} \quad \hat U(0) = 1 .
\la{eq:SE}
\ee
In particular, for constant $h$
\be
\nu(t) = h t \quad \hbox{and} \quad \tau(t) = \frac{1- e^{-ht}}{h}.
\la{eq:newtimes}
\ee
The evolution of a massless bosonic wave packet in the expanding space can therefore be represented as the composition of the propagation in stationary space, albeit with slowed-down time $\tau(t)$ of \Eq{eq:newtime} (called ``conformal time'' in the cosmological literature), and the dilation map.  For a plane-wave mode with $\hat H | \vv k \rangle = | \vv k | \cdot | \vv k \rangle$, the condition $\tau (t) \cdot | \vv k | \ll 1$ implies that the time-evolution operator of \Eq{eq:U-wave} is completely dominated by the dilation $\exp{[-i \nu(t) \hat D]}$.  This corresponds to the ``freezing in'' of modes with wavelengths larger than the causal horizon (``super-horizon modes'') during inflation.  Note that throughout this article we work in terms of the physical (rather than the comoving) momenta.

%%%%%%%%%%
%%% DE SITTER VACUUM AS HEAT BATH
%%%%%%%%%%

\subsection{de Sitter vacuum as heat bath}
\la{sec:dS-bath}

We consider a dS universe in the vacuum state $| \Omega \rangle$ of the massless scalar field.  The physics of this state can be tested by putting a localized system, such as the harmonic oscillator of \Eq{eq:Hosc}, in contact with the background scalar field.  This will then serve as a thermometer (see discussion at the end of \Sec{sec:spectral}).  The quantum noise acting on this system is fully characterized by the background field's second order correlation function, or by its Fourier transform [\Eq{eq:spectral}].  In order to characterize the properties of the dS vacuum we compute the analog of the bath correlation function $G_\Lambda(t)$ that appeared in \Eq{eq:spectral}:
\be
G^{\rm dS}_\Lambda(t) = \langle\Omega| e^{i\hat H_{\rm dS} t}\hat{\phi}_{\Lambda}(0)e^{-i\hat H_{\rm dS} t}\hat{\phi}_{\Lambda} (0) |\Omega\rangle
\la{eq: correlation}
\ee
where $\hat{\phi}_{\Lambda}(0)$ is given by \Eq{eq:regfield} with $\omega(\vv k) = |\vv k|$, and $\hat H_{\rm dS}$ is the second quantization (i.e., lifting to Fock space) of the single-boson Hamiltonian $\hat H_D$ in \Eq{eq:H1a} for $h =$ const.  Then
\be
G^{\rm dS}_{\Lambda}(t)  =  \langle g_{\Lambda} | \hat U (t) | g_{\Lambda} \rangle ,
\la{eq:Gt}
\ee
where $\hat U(t)$ is given by Eqs.\ \eqref{eq:U-wave}, \eqref{eq:newtimes}, and the momentum representation of the wave function for $g_{\Lambda}$,
\be 
\tilde \psi (\vv k) = \langle \vv k | g_{\Lambda} \rangle = \frac{e^{-|\vv k|/\Lambda}}{\sqrt{2 | \vv k |}} .
\la{eq:glambda}
\ee
The corresponding wave function in position representation is
\be
\psi (\vv x) = \langle \vv x | g_{\Lambda} \rangle = \int d^3 k \, \langle \vv x | \vv k \rangle \langle \vv k | g \rangle
=  \int \frac{d^3 k}{(2 \pi)^{3/2}} \frac{e^{- i \vv k \cdot \vv x} e^{- | \vv k | / \Lambda}}{\sqrt{2|\vv k|}} ,
\la{eq:gposition}
\ee
in the convention for the Fourier transform in which
\be
\langle \vv x | \vv k \rangle = \frac{e^{- i \vv k \cdot \vv x}}{(2\pi)^{3/2}} .
\la{eq:Fourier}
\ee

In this case,
\be
\langle \vv x | e^{-i \tau (t) \hat H} | g_{\Lambda} \rangle
= \int d^3 k \, \langle \vv x | e^{-i \tau(t) \hat H} | \vv k \rangle \langle \vv k | g_{\Lambda} \rangle
= \int d^3 k \, e^{-i | \vv k | \tau(t)} \langle \vv x | \vv k \rangle \langle \vv k | g \rangle
= \int \frac{d^3 k}{(2 \pi)^{3/2}} \frac{e^{-i | \vv k | \tau (t)} e^{-i \vv k \cdot \vv x} e^{- | \vv k | / \Lambda} }{\sqrt{2|\vv k|}} .
\ee
Using \Eq{eq:D} for the action of the dilation operator on the wave function, we have that
\begin{align}
\langle \vv x | e^{i \nu (t) \hat D} | g_{\Lambda} \rangle &= \left[ e^{i \nu(t) \hat D} \psi \right] (\vv x) = \frac{e^{\frac 3 2 \nu (t)}}{(2 \pi)^{3/2}} \int d^3 k' \, \frac{e^{- | \vv k' | / \Lambda}}{\sqrt{2|\vv k'|}} e^{-i \vv k' \cdot \vv x e^{\nu(t)}} .
\end{align}
The bath correlation function of \Eq{eq:Gt} can therefore be expressed as
\begin{align}
G^{\rm dS}_{\Lambda}(t) &= \int d^3 x \, \langle g_\Lambda | e^{-i \nu (t) \hat D} | \vv x \rangle \langle \vv x | e^{-i \tau (t) \hat H} | g_\Lambda \rangle
= \frac{1}{(2 \pi)^3} \int d^3 x \, d^3 k \, d^3 k' \, e^{\frac 3 2 \nu (t)}  \frac{e^{- ( | \vv k | + | \vv k' | ) / \Lambda}}{2\sqrt{|\vv k| \cdot |\vv k'|}} e^{- i | \vv k | \tau (t)} e^{-i ( \vv k - e^{\nu(t)} \vv k') \cdot \vv x} \nl
&= e^{\frac 3 2 \nu (t)} \int d^3 k \, d^3 k' \, \frac{e^{- ( | \vv k | + | \vv k' | ) / \Lambda}}{2 \sqrt{|\vv k| \cdot |\vv k'|}}  e^{- i | \vv k | \tau (t)} \delta^{(3)} \left( \vv k - e^{\nu (t)} \vv k' \right)
= e^{- \frac 3 2 \nu(t)} \int d^3 k \,\frac{ e^{- \left( 1 + e^{-\nu(t)} \right) | \vv k | / \Lambda}}{2|\vv k| e^{- \nu(t)/2}} e^{- i |\vv k| \tau (t)} \nl
&= 2 \pi e^{- \nu(t)} \int_0^\infty dr \, r \exp{\left[ - r \left( \frac{1 + e^{-\nu(t)}}{\Lambda} + i \tau (t) \right) \right]}
= 2 \pi e^{- \nu(t)} \left( \frac{1 + e^{-\nu(t)}}{\Lambda} + i \tau (t) \right)^{-2} .
\end{align}
For constant $h$ this becomes
\be
G^{\rm dS}_{\Lambda}(t) = \frac \pi 2 \left( \frac{h \Lambda}{h \cosh \frac{h t}{2} + i \Lambda \sinh \frac{h t}{2}} \right)^2 .
\la{eq:GdS}
\ee

In the convention of \Eq{eq:spectral}, the corresponding spectral density is
\be
\tilde G^{\rm dS}_{\Lambda} (\omega) = 2 \pi^2 \Lambda^2 \frac{\omega}{h^2 + \Lambda^2} \hbox{csch}{\left( \frac{\pi \omega}{h} \right)} \exp{\left[ \frac{2 \omega}{h} \arcsin \left( \frac{\Lambda}{\sqrt{h^2 + \Lambda^2}} \right) \right]} .
\la{eq:tGdS}
\ee
Note that $\tilde G^{\rm dS}_{\Lambda} (\omega) > 0$, as required by Bochner's theorem (see also the discussion in the Appendix).  The fact that this spectral density has support on $\omega < 0$ indicates a dynamical instability associated with particle production.  We can compute the Boltzmann factor according to \Eq{eq:KMS}, which gives us
\be
\beta = \frac 4 h \arcsin \left( \frac{\Lambda}{\sqrt{\Lambda^2 + h^2}} \right) .
\la{eq:beta}
\ee
In the limit $\Lambda \to \infty$, this gives the usual dS temperature
\be
T_{\rm dS} = \frac 1 \beta = \frac{h}{2\pi} .
\la{eq:TdS}
\ee
The dS spectral density is
\be
\tilde G^{\rm dS} (\omega) \equiv \lim_{\Lambda \to \infty} \tilde G^{\rm dS}_{\Lambda} (\omega)
= 2 \pi^2 \omega \, \hbox{csch} \left(\frac{\omega}{2 T_{\rm dS}} \right) e^{\omega / 2T_{\rm dS}}
= \frac{(2 \pi)^2 \omega}{1 - e^{-\omega / T_{\rm dS}}} .
\la{eq:tGdS1}
\ee
Comparing this to \Eq{eq:spectralfin}, we conclude that dS space acts as a thermal bath with density of states
\be
n^{\rm dS}(\omega) = \frac{ \omega \tilde G^{\rm dS} (\omega) \left( 1 - e^{- \omega / T_{\rm dS}} \right)}{8 \pi^4} =  \frac{\omega^2}{2 \pi^2} ,
\la{eq:ndS}
\ee
as in \Eq{eq:nPl}.

The energy density of the dS bath obeys the Stefan-Boltzmann law:
\be
\rho_{\rm dS} = g_f \int_0^\infty d \omega \, {\cal U}^{\rm dS} (\omega)
= g_f \int_0^\infty d\omega \, \frac {n^{\rm dS}(\omega)\, \omega}{e^{\omega / T_{\rm dS}} - 1}
= \frac{g_f}{2 \pi^2} \int_0^\infty \frac{d \omega \, \omega^3}{e^{\omega / T_{\rm dS}} - 1}
= \frac{g_f T_{\rm dS}^4}{2 \pi^2} \int_0^\infty \frac{dx \, x^3}{e^x - 1}
= \frac{g_f \pi^2}{30} T_{\rm dS}^4 ,
\la{eq:SB}
\ee
where $g_f$ denotes the number of degrees of freedom in the bath (which we have assumed to correspond to massless fields), and where we have used the energy density spectrum of \Eq{eq:energydensity} with the density of states of \Eq{eq:ndS}.  In terms of the Hubble parameter,
\be
\rho_{\rm dS} = \sigma h^4 , \quad \hbox{for} \quad \sigma \equiv \frac{g_f}{480 \pi^2} .
\la{eq:sigma}
\ee

It should be emphasized that, even though in the discussion of \Sec{sec:MME} we took the open system in the MME [\Eq{eq:MME}] to be a simple harmonic oscillator, our results do not depend on the details of that system, which can be thought of as a generic quantum thermometer.  Equation \eqref{eq:tGdS1} implies that {\it any} local system with a discrete spectrum that is weakly coupled to the massless scalar field in dS space will thermalize at temperature $T_{\rm dS}$.  If the field has a small self-coupling, then the modes of the field will themselves thermalize to that temperature.  This implies that a cosmological dS phase (inflation) is accompanied by particle pair production, which in equilibrium gives the constant energy density of \Eq{eq:sigma}, assuming that the $g_f$ polarizations are all massless.

Note also that, in line with the arguments of \cite{Sewell} for physical heat baths, only the observer in the cosmic rest frame will measure $T_{\rm dS}$.  Boosts with respect to this frame break the KMS condition of \Eq{eq:KMS} due to Doppler shifts of the mode frequencies, so that the bath will not appear in equilibrium (the results of \cite{rotating, tribo}, in which Doppler shifts were introduced by rotational motion, can be easily extended to linear motion).  This is an important difference with the original results of Gibbons and Hawking \cite{GH}, in which $T_{\rm dS}$ was interpreted as preserving the isometries of dS space.  Polyakov has emphasized in \cite{Polyakov12} that the absence of such Doppler shifts prevents us from interpreting the result of \cite{GH} in terms of a physical heat bath.

We calculated $\tilde G^{\rm dS} (\omega)$ and the associated thermal properties for a single background massless scalar field.  In fact, all of the quantum fields present should be included in the calculation of the bath correlation function.  Fortunately, it is not difficult to extend our results in that way, as long as $h$ is large enough that the masses of the fields can be neglected.  For fermions, the thermal power spectrum and the energy density spectrum [Eqs.\ \eqref{eq:Pdensity2} and \eqref{eq:energydensity}] become, respectively,
\be
\mathcal{P}_{\rm th}(\omega)= (2\pi)^3\frac{n(\omega)}{\omega(e^{\beta\omega} + 1)} \quad \hbox{and} \quad
\mathcal{U}(\omega) = \frac {n(\omega)\, \omega}{e^{\beta\omega} + 1} .
\la{eq:fermionspectra}
\ee
For $\beta \omega \ll 1$, the contribution of fermions to the thermal power spectrum can therefore be neglected relative to the contribution from bosons.  The fermion contribution to $\rho_{\rm dS}$ in \Eq{eq:SB} is weighed by a factor of
\be
\left. \int_0^\infty \frac{dx \, x^3}{e^x + 1} \right/ \int_0^\infty \frac{dx \, x^3}{e^x - 1} = \frac 7 8 ,
\la{eq:gf}
\ee
as is well known in cosmology for the energy density of ordinary thermal radiation (see, e.g., Sec.\ 3.3 in \cite{Kolb-Turner}).

%%%%%%%%%%
%%% INFRARED INSTABILITY
%%%%%%%%%%

\subsection{Infrared instability}
\la{sec:IR}

The results of \Sec{sec:dS-bath} were derived assuming $h =$  const.  As long as the space-time geometry remains homogenous and isotropic, and if the change in $h$ is very slow compared to the Planck-time scale that controls the gravitational microphysics, we may use \Eq{eq:sigma} as an adiabatic relation between $h(t)$ and the heat bath's contribution $\rho_{\rm dS} (t)$ to the energy density of the early Universe.  By including terms corresponding to this contribution in the Friedmann Eqs.\ \eqref{eq:1Fried} and \eqref{eq:2Fried}, we can take into account the back-reaction of the heat bath on the space-time.  This is done in detail in the accompanying letter \cite{relax}.

The bath with energy density $\rho_{\rm dS}$ can be considered as a cosmological fluid whose pressure $p_{\rm dS}$ obeys the equation of state $w_{\rm dS} = p_{\rm dS} / \rho_{\rm dS} = -1$.  The bath must obey this equation of state because the energy density of \Eq{eq:SB} is not diluted by the Universe's expansion, as per \Eq{eq:conserve} (the increasing amount of energy in this fluid as the Universe expands is compensated by the negative energy of gravity).  We may therefore write the total energy density of the Universe as
\be
\rho = \rho_{\rm dS} + \rho_r = \sigma h^4 + \rho_r
\la{eq:rhos}
\ee
and the total pressure as
\be
p = p_{\rm dS} + p_r = - \sigma h^4 + w_r \rho_r .
\la{eq:ps}
\ee
By combining Eqs.\ \eqref{eq:rhos} and \eqref{eq:ps} with the Friedmann equations, we show in \cite{relax} that ---much like in a superheated liquid--- a perturbation may trigger a spontaneous boiling, which in the cosmological case corresponds to the production of radiation with equation of state $w_r =  p_r / \rho_r = 1/3$.  This boiling relaxes the temperature, so that $h \to 0$.  This leads us to conclude that dS space is, as previous authors have argued on a variety of grounds, unstable in the IR.  This picture gives us the main features of inflation without invoking any inflaton potential function, or even a coherent inflaton field.\footnote{In \cite{relax} we find that such inflation will naturally occur at the Planck scale (i.e., \hbox{$T_{\rm dS} \sim M_{\rm Pl}$}).  In that case the thermodynamic treatment is well justified, because there is a very large separation of scales between the energy density of the dS vacuum and the local dynamics of any Standard Model probe.}

In the case of boiling water, the microscopic details are complicated, but the process can be understood macroscopically with the aid of the laws of thermodynamics.  A microscopic description of inflation would require a better understanding of quantum gravity than what is currently available, but the methods of quantum thermodynamics, combined with the classical Friedman equations for the macroscopic space-time, allow us to describe such an irreversible relaxation in broad strokes.  On the question of the size of the initial perturbation that triggers the relaxation, \cite{relax} invokes another analogy to a phase transition: the phenomenon of Dicke superradiance in quantum optics \cite{Dicke, Gross-Haroche}.

The interplay between local pair production (which is a UV effect) and the decay of the dS vacuum (which is an IR effect) may be related to the ``IR/UV mixing'' that Polyakov discusses in \cite{Polyakov12}.  Note that, in contrast to our local approach, in \cite{Mirbabayi} Mirbabayi considers only the relaxation of extended (delocalized) IR modes.  It is therefore not surprising, from our perspective, that he should find dS to be stable, since that description does not incorporate the local pair production.

Despite our limited understanding of quantum gravity, our results may have something to say about point (a) in the list given in \Sec{sec:intro}:  the problem of the very small entropy of the initial state of the Universe.  Our picture is consistent with taking the initial state of the Universe to be the Bunch-Davies vacuum \cite{BD} with $\rho_{\rm dS} \sim M_{\rm Pl}^4$ (the ``large cosmological constant'').  An observer with access only to a subspace of the full Hilbert space of the Universe may see that vacuum state as having positive entropy.  The irreversible relaxation of inflation ($\rho_{\rm dS} \to 0$) described in \cite{relax} could therefore be compatible with a pure initial state.  That relaxation may be thought of as a tunneling between the initial Bunch-Davies vacuum and a final Minkowski vacuum.  Both states are devoid of particles so that, if the negative energy contribution of gravity is included in the total energy, no global conservation (``superselection'') prevents such a tunneling.

According to the astronomical evidence first reported in \cite{Riess, Perlmutter}, the Universe does not appear to be relaxing steadily towards empty Minkowski space any more, but has instead recently entered a new dS phase in which matter and radiation coexist with $\rho_{\rm dS} \sim 10^{-123} M_{\rm Pl}^4$.  Our work, unfortunately, has nothing to say about this ``small cosmological constant'' problem.  It might be due to the presence of a condensate with equation of state $w = -1$, fated to evaporate due to the instability of dS space.  Perhaps in a more complete theory of quantum gravity the adiabatic relation between the bath energy density $\rho_{\rm dS}$ to the Hubble parameter $h$, used in \cite{relax}, might break down at some time after the end of inflation, obstructing the further relaxation of $\rho_{\rm dS}$.  Note also that in all our calculations the background fields were taken to be massless.  This is a good approximation for the early Universe, when $h$ was close to the Planck scale, but the effect of particle masses on the thermodynamics of a dS space at the current, much smaller value of $h$ requires further investigation.
 
%%%%%%%%%%
%%% DISCUSSION AND OUTLOOK
%%%%%%%%%%

\section{Discussion and outlook}
\la{sec:discussion}

In this paper we have applied the analytical machinery of quantum thermodynamics, based on the MME for an open system, to the local physics as seen by an observer in dS space whose space-time metric takes the FLRW form of \Eq{eq:FLRW} (the ``cosmic rest frame'').  Though widely used in quantum optics and some other areas of physics, such analytical techniques have, to our knowledge, not been applied before in the context of dS thermodynamics.  This may be partly because they require a noncovariant formulation, which can seem problematic from the point of view of GR.  However, because thermodynamic quantities are noncovariant \cite{Sewell} and because cosmology is principally concerned with the physics as it appears to an observer in the cosmic rest frame, we regard this as a useful feature of the MME approach.

We showed that an observer in the cosmic rest frame will see a local system weakly coupled to a background massless quantum field as subject to a black-body radiation with temperature $T_{\rm dS} = h / 2 \pi$.  Unlike in \cite{GH}, where this temperature was first computed, we find that other inertial observers do not measure the same $T_{\rm dS}$, but instead see the radiation as out of equilibrium, because the Doppler shifts of the mode frequencies break the KMS condition of \Eq{eq:KMS}.  This allows us to interpret $T_{\rm dS}$ as the temperature of a {\it physical} bath with an energy density obeying the Stefan-Boltzmann law of \Eq{eq:SB}.  In terms of the interpretation that the authors of \cite{trouble} attribute to Shenker, this reflects an {\it anomaly}: the quantum theory of gravity does not respect the classical symmetries of dS space.

If the Stefan-Boltzmann law ($\rho_{\rm dS} \propto h^4$) is extended adiabatically to the case of slowly decreasing $h$, the backreaction of $\rho_{\rm dS}$ on the space-time may be taken into account via the Friedmann equations from classical GR.  In that case, we find that $h$ relaxes irreversibly towards zero via particle production.  This supports previous arguments for the IR instability of dS space, such as those offered in \cite{Myhrvold, Ford-IR, Mottola, Antoniadis, Tsamis-Woodard96, Tsamis-Woodard98, Polyakov07, Polyakov12, Dvali-Gomez, Akhmedov}, while clearing up some of the obscurity that has surrounded the thermodynamics of dS space.  The implications of these results for cosmology are covered in \cite{relax}.

In \cite{relax} we also propose that the primordial perturbations that seed structure formation in the Universe, generally interpreted as resulting form the ``freeze-in'' of vacuum fluctuations, may be interpreted instead as arising from the thermal fluctuations described by the power spectrum ${\cal P}_{\rm th} (\omega)$ in \Eq{eq:Pdensity2}.  This proposal is akin to that of ``warm inflation'' (see \cite{warm1, warm2} and references therein), except that in our case the temperature of the fluctuations is given directly by the $h$ of inflation and that the thermal fluctuations are obtained from the quantum noise spectra that we introduced in \Sec{sec:noise}, rather than from a classical Langevin equation.  Moreover, our picture does not invoke an inflaton potential $V(\phi)$, or even a coherent inflaton field $\phi$.  It is clear that such perturbations will be adiabatic, Gaussian, and approximately scale invariant, but detailed calculation of cosmological observables in such a qualitatively new implementation of inflation requires further investigation and the development of a calculational framework not based on a coherent inflaton.

In addition to clarifying the thermodynamics of dS space and inflation, we believe that the MME framework developed in this article can offer a new theoretical tool for understanding the history of the early Universe in terms of out-of-equilibrium, irreversible dynamics.  This irreversibility entirely breaks the (approximate) time-reversal symmetry of the underlying microphysics.  Cohen and Kaplan have shown that the breaking of the (microphysically exact) $CPT$ invariance by the time-dependent background during inflation allows baryogenesis without $CP$ violation and while baryon-violating interactions are still in equilibrium, circumventing the Sakharov conditions \cite{Cohen-Kaplan}.  We believe that the problem of baryogenesis should, therefore, be reexamined in light of the irreversible dynamics of the early Universe. \\

%%%%%%%%%%
%%% ACKNOWLEDGMENTS
%%%%%%%%%%

{\bf Acknowledgements}: We thank Jonathan Oppenheim, Enrico Pajer, and Fernando Quevedo for discussions.  RA and AJ were supported by the International Research Agendas Programme (IRAP) of the Foundation for Polish Science (FNP), with structural funds from the European Union (EU).  GB was supported by the Spanish Frants No.\ PID2020-113334GB-I00/AEI/1013039/501100011033 and No.\ CIPROM/2021/054 (Generalitat Valenciana). GB has also received support from the EU's Horizon 2020 research and innovation program under the Marie Sk{\l}odowska-Curie Grant Agreement No.\ 860881-HIDDeN.

%%%%%%%%%%
%%% REFERENCES
%%%%%%%%%%

\newpage

%%%%%%%%%%
%%% APPENDIX
%%%%%%%%%%

\appendix

\section{Open-system formalism}
\la{sec:appendix}

For the convenience of readers who might not be familiar with the analytical methods for open quantum systems as covered, e.g., in \cite{Alicki-Lendi}, we include in this Appendix a brief summary of salient points of the general derivation of the MME as they are relevant to the present work.  A similar but slightly more extensive summary is given in the appendix of \cite{rotating}.  For an interesting review of the history of the development of this formalism, see \cite{historyGKLS}.  As we pointed out at the beginning of \Sec{sec:MME}, the MME has been previously applied to cosmology to describe pure decoherence effects (see, e.g., \cite{Colas-etal} and references therein), whereas we are interested in using it to describe the thermodynamics of dS space.

The MME describes the nonunitary evolution of the reduced density matrix $\hat \rho (t)$ of an open system $S$ weakly coupled to a bath $B$ in stationary state $\hat \rho_B$. The system and bath Hamiltonians are denoted by $\hat H_0$ and $\hat H_B$, respectively.  The interaction Hamiltonian is taken to be the product of two Hermitian operators
\be
\lambda \hat H_{\rm int} = \lambda \hat S \otimes \hat B ,
\la{eq:hamint0}
\ee
 with the constant $\lambda$ giving the strength of the coupling.  We may always assume that $\Tr \left( \hat \rho_B \, \hat B \right) = 0$.  The reduced dynamics of $S$ is expressed in the interaction picture:
\be
\hat\rho (t) = \Lambda (t,0) \hat\rho \equiv \Tr_B \left[ \hat U_\lambda (t,0) \hat\rho \otimes \hat\rho_B \hat U_\lambda (t,0)^\dagger \right] ,
\la{eq:red_dyn}
\ee
for
\be
\hat U_\lambda (t,0) = {\mathbf T} \exp \left[ -i \lambda \int_0^t ds \, \hat S(s) \otimes \hat B(s) \right] ,
\la{eq:prop_int}
\ee
(where the symbol $\mathbf T$ indicates time ordering), with
\be
\hat S(t) = e^{i \hat H t} \hat S e^{- i \hat H t} \quad \hbox{and} \quad  \hat B(t)= e^{i \hat H_B t} \hat B e^{- i \hat H_B t} .
\la{eq:prop_int1}
\ee
Here, $\hat S(t)$ is defined in terms of the {\it renormalized} Hamiltonian 
\be
\hat H = \hat H_0 + \lambda^2 \hat H_{1}^{\rm corr} + \ldots  
\la{eq:H_S}
\ee
The terms proportional to powers of $\lambda $ in \Eq{eq:H_S} are Lamb-shift corrections due to the interaction with the bath, which cancel with the uncompensated term $\hat H - \hat H_0$ that should, in principle, be present in \Eq{eq:prop_int}. 
\par
To extract the leading contribution to the reduced dynamics we apply the cumulant expansion
\be
\Lambda (t,0) = \exp \sum_{n=1}^\infty \left[ \lambda^n {\cal K}^{(n)} (t) \right]
\la{eq:cumulant} .
\ee
The first term is ${\cal K}^{(1)} = 0$ and the leading Born approximation (corresponding to weak coupling) corresponds to taking ${\cal K} = {\cal K}^{(2)}$, so that 
\be
\Lambda (t,0) = \exp \left[ \lambda^2 {\cal K}(t) + O \left( \lambda^3 \right) \right] .
\ee
By direct comparison of the cumulant expansion of \Eq{eq:cumulant} and Dyson expansion for \Eq{eq:red_dyn} we get that
\be
{\cal K}(t) \hat\rho = \int_0^t ds \int_0^t du \, G(s-u) \hat S(s) \hat\rho \hat S^\dagger (u) +(\rm{similar\ terms}) ,  
\la{eq:K(t)}
\ee
where the {\it bath correlation function} is
\be
G(s) \equiv \Tr \left[ \hat \rho_B \hat B(s) \hat B \right] .
\la{eq:bcf}
\ee
The ``similar terms'' in \Eq{eq:K(t)} are of the form $\hat\rho \hat S(s) \hat S^\dagger (u)$ and $\hat S(s) \hat S^\dagger (u) \hat\rho$.

In the interaction picture, the Markovian approximation, which is valid for sufficiently long times $t$, is
\be
{\cal K}(t) \simeq t {\cal L} ,
\la{eq:L}
\ee
where $\cal L$ is the Gorini-Kossakowski-Lindblad-Sudarshan (GKLS) generator \cite{GKS, Lindblad}.  We decompose  $S(t)$  into its Fourier components
\be
\hat S(t) = \sum_{\{\omega\} } e^{i\omega t} \hat S_\omega , ~~\hbox{with}~~  \hat S_{-\omega } = \hat S_\omega^\dagger ,
\la{eq:S}
\ee
where the set $\{\omega\}$ contains the ``Bohr frequencies'' of the Hamiltonian:
\be
\hat H = \sum_k \epsilon_k | k \rangle \langle k |, ~~  \omega = \epsilon_k - \epsilon_l .  
\la{Bohr}
\ee
This allows us to write \Eq{eq:K(t)} in the form
\be
{\cal K}(t) \rho = \sum_{\omega ,\omega'} \hat S_\omega \hat\rho \hat S_{\omega'}^\dagger \int_0^t du\, e^{i(\omega -\omega') u} \int_{-u}^{t-u} d\tau \, G(\tau) e^{i\omega \tau} + (\rm{similar\ terms}) .
\la{eq:K2}
\ee
Applying the approximations
\be
\int_0^t du \, e^{i(\omega -\omega')u} \approx t \delta _{\omega \omega'} ~~\hbox{and}~~
\int_{-u}^{t-u} d \tau \, G(\tau) e^{i\omega \tau} \approx \tilde G (\omega) \equiv \int_{-\infty}^\infty d\tau \, G(\tau) e^{i\omega \tau} \geq 0 ,
\la{eq:rep1}
\ee
which make sense for $t \gg \max \{1/(\omega -\omega')\}$, we obtain that
\be
{\cal K}(t) \hat\rho = t \sum_\omega \tilde G(\omega)\hat S_\omega \hat\rho \hat S_\omega^\dagger  + (\hbox{similar\ terms})\equiv t{\cal L} ,
\ee
where $\tilde G(\omega)$ is called the bath's {\it spectral density}.  Using the commutation of ${\cal L}$ with $[H , \cdot]$ one can easily return to the Schr\"odinger picture and obtain the final MME: 
\bea
\frac{d \hat\rho}{dt} &=& - i \left[ \hat H,\hat \rho \right] + {\cal L} \hat \rho , \notag \\
{\cal L} \hat\rho &\equiv& \frac{\lambda^2}{2} \sum_{\{\omega \}} \tilde G(\omega) \left( \left[ \hat S_\omega , \hat\rho \hat S_\omega^\dagger \right] + \left[ \hat S_\omega \hat\rho , \hat S_\omega^\dagger \right] \right) .  
\la{eq:Dav}
\eea

The following points are important to bear in mind:

\begin{enumerate}[label=(\roman*)]

\item The absence of off-diagonal terms in \Eq{eq:Dav}, compared to \Eq{eq:K2}, is the crucial property, which can be interpreted as a coarse-graining in time of rapidly oscillating terms (the ``secular approximation''). It implies the commutation of $\cal L$ with the Hamiltonian part $[\hat H, \cdot]$.

\item The positivity $\tilde G(\omega) \geq 0$ follows from Bochner's theorem and is a necessary condition for \emph{complete positivity} of the MME, which is a fundamental property of quantum dynamics that prevents the occurrence of negative probabilities.  When applying this formalism to a particular open system, this positivity of the bath's spectral density can provide a useful sanity check: see, e.g., \Eq{eq:tGdS} in the main text.

\item For more complicated interaction Hamiltonians of the form $\hat H_{\rm int} = \sum_\alpha \hat S_\alpha \otimes \hat B_\alpha$ we replace $\tilde G(\omega)$ by the positive-definite relaxation matrix $\tilde G_{\alpha \beta}(\omega)$, which (because of symmetry) is usually diagonal in an appropriate parametrization.

\item Point (i) implies that the diagonal elements of the density matrix (in the energy representation) evolve independently of the off-diagonal ones. They satisfy the Pauli master equation, with transition rates equal to those calculated from Fermi's golden rule \cite{golden}.

\end{enumerate}

If the reservoir is a quantum system in a thermal equilibrium state, then the bath's spectral density must obey the KMS condition
\be
e^{- \beta \omega} = \frac{\tilde G(-\omega)}{\tilde G(\omega)} ,  
\la{eq:KMS1}
\ee
where $\beta$ is the bath's inverse temperature ($\beta = 1 / T$).  As a consequence of \Eq{eq:KMS1}, the Gibbs state 
\be
\hat\rho_\beta = Z^{-1} e^{-\beta \hat H},
\la{eq:Gibbs}
\ee
is a stationary solution of \Eq{eq:Dav}. Under mild conditions (e.g., that the only system operators commuting with $\hat H$ and $\hat S$ are scalars) the Gibbs state is a unique stationary state and any initial state relaxes towards equilibrium (``zeroth law of thermodynamics''). One can also show that the second law of thermodynamics is fulfilled in the sense that entropy production is positive.

Note that if an observer sees the bath moving, the Doppler shifts of the $\omega$'s in \Eq{eq:KMS1} will break the KMS condition, so that the bath will not fulfill the zeroth law of thermodynamics with respect to that observer.  Sewell showed in \cite{Sewell} that temperature cannot, therefore, be defined covariantly in quantum physics.  Some of the physical consequences of such Doppler shifts have been studied in \cite{rotating, tribo}.


\begin{thebibliography}{99}

%%%%%%%%%%
%%% INTRODUCTION
%%%%%%%%%%

\bibitem{Hubble}
	E.~Hubble,
	``A relation between distance and radial velocity among extra-galactic nebulae'',
	Proc.\ Natl.\ Acad.\ Sci.\ USA \href{https://doi.org/10.1073/pnas.15.3.168}{{\bf 15}, 168} (1929) 

\bibitem{Schroedinger}
	E.~Schr\"odinger,
	``The proper vibrations of the expanding Universe'',
	Physica \href{https://doi.org/10.1016/S0031-8914(39)90091-1}{{\bf 6}, 899} (1939)
	
\bibitem{Ford-rev}
	L.~H.~Ford,
	``Cosmological particle production: A review'',
	Rep.\ Prog.\ Phys.\ \href{https://doi.org/10.1088/1361-6633/ac1b23}{{\bf 84}, 116901} (2021)
	[arXiv:2112.02444 [gr-qc]]
	
\bibitem{GH}
	G.~W.~Gibbons and S.~W.~Hawking,
	``Cosmological event horizons, thermodynamics, and particle creation'',
	Phys.\ Rev.\ D \href{https://doi.org/10.1103/PhysRevD.15.2738}{{\bf 15}, 2738} (1977)
	
\bibitem{LesHouches}
	M.~Spradlin, A.~Strominger, and A.~Volovich,
	``Les Houches lectures on de Sitter space'', 
	arXiv:hep-th/0110007
	
\bibitem{Anninos}
	D.~Anninos,
	``De Sitter musings'',
	Int.\ J.\ Mod.\ Phys.\ A \href{https://doi.org/10.1142/S0217751X1230013X}{{\bf 27}, 1230013} (2012)
	[arXiv:1205.3855 [hep-th]]
	
\bibitem{deAlwis}
	S.~P.~de Alwis,
	``Comments on entropy calculations in gravitational systems'',
	arXiv:2304.07885 [hep-th]
	
\bibitem{Guth}
	A.~H.~Guth,
	``Inflationary universe: A possible solution to the horizon and flatness problems'',
	Phys.\ Rev.\ D \href{https://doi.org/10.1103/PhysRevD.23.347}{{\bf 23}, 347} (1981)
	
\bibitem{Weinberg}
	S.~Weinberg,
	{\it Cosmology}
	(Oxford: Oxford University Press, 2008)
	
\bibitem{Riess}
	A.~G.~Riess et al.\ [Supernova Search Team Collaboration],
	``Observational evidence from supernovae for an accelerating universe and a cosmological constant,''
	Astron.\ J.\  {\bf 116}, 1009 (1998)
	[arXiv:astro-ph/9805201]
	
\bibitem{Perlmutter}
	S.~Perlmutter et al.\ [Supernova Cosmology Project Collaboration],
	``Measurements of omega and lambda from 42 high-redshift supernovae,''
	Astrophys.\ J.\  {\bf 517}, 565 (1999)
	[arXiv:astro-ph/9812133]
	
\bibitem{GKS}
	V.~Gorini, A.~Kossakowski, and E.~C.~G.~Sudarshan,
	``Completely positive dynamical semigroups of $N$-level systems'',
	J.\ Math.\ Phys.\ (NY) \href{https://doi.org/10.1063/1.522979}{{\bf 17}, 821} (1976)
	
\bibitem{Lindblad}
	G.~Lindblad,
	``On the generators of quantum dynamical semigroups'',
	Commun.\ Math.\ Phys.\ \href{https://doi.org/10.1007/BF01608499}{{\bf 48}, 119} (1976)
	
\bibitem{LL-CRF}
	A.~R.~Liddle and D.~H.~Lyth,
	{\it Cosmological Inflation and Large-Scale Structure}
	(Cambridge: Cambridge University Press, 2000), Sec.\ 4.4
	
\bibitem{relax}
	R.~Alicki, G.~Barenboim, and A.~Jenkins,
	``The irreversible relaxation of inflation'',
	arXiv:2307.04803 [gr-qc]
	
\bibitem{Myhrvold}
	N.~Myhrvold,
	``Runaway particle production in de Sitter space'',
	Phys.\ Rev.\ D \href{https://doi.org/10.1103/PhysRevD.28.2439}{{\bf 28}, 2439} (1983)
	
\bibitem{Ford-IR}
	L.~H.~Ford,
	``Quantum instability of de Sitter spacetime'',
	Phys.\ Rev.\ D \href{https://doi.org/10.1103/PhysRevD.31.710}{{\bf 31}, 710} (1985)
	
\bibitem{Mottola}
	E.~Mottola,
	``Particle creation in de Sitter space'',
	Phys.\ Rev.\ D \href{https://doi.org/10.1103/PhysRevD.31.754}{{\bf 31}, 754} (1985)
	
\bibitem{Antoniadis}
	I.~Antoniadis, J.~Iliopoulos, and T.~N.~Tomaras,
	``Quantum instability of de Sitter space'',
	Phys.\ Rev.\ Lett.\ \href{https://doi.org/10.1103/PhysRevLett.56.1319}{{\bf 56}, 1319} (1986)
	
\bibitem{Tsamis-Woodard96}
	N.~C.~Tsamis and R.~P.~Woodard,
	``Quantum gravity slows inflation'',
	Nucl.\ Phys.\ B \href{https://doi.org/10.1016/0550-3213(96)00246-5}{{\bf 474} 235} (1996)
	[arXiv:hep-ph/9602315]
	
\bibitem{Tsamis-Woodard98}
	N.~C.~Tsamis and R.~P.~Woodard,
	``Non-perturbative models for the quantum gravitational back-reaction on inflation'',
	Ann.\ Phys.\ (NY) \href{https://doi.org/10.1006/aphy.1998.5816}{{\bf 267}, 145} (1998)
	[arXiv:hep-ph/9712331]
	
\bibitem{Polyakov07}
	A.~M.~Polyakov,
	``De Sitter space and eternity'',
	Nucl.\ Phys.\ B \href{https://doi.org/10.1016/j.nuclphysb.2008.01.002}{{\bf 797}, 199} (2008)
	[arXiv:0709.2899 [hep-th]]

\bibitem{Polyakov12}
	A.~M.~Polyakov,
	``Infrared instability of the de Sitter space''
	arXiv:1209.4135 [hep-th]
	
\bibitem{Dvali-Gomez}
	G.~Dvali and C.~Gomez,
	``Quantum compositeness of gravity: Black holes, AdS and inflation'',
	J.\ Cosmol.\ Astropart.\ Phys.\ \href{https://doi.org/10.1088/1475-7516/2014/01/023}{{\bf 1401}, 023} (2014)
	[arXiv:1312.4795 [hep-th]]
	
\bibitem{Akhmedov}
	E.~T.~Akhmedov, U.~Moschella, and F.~K.~Popov,
	``Characters of different secular effects in various patches of de Sitter space'',
	Phys.\ Rev.\ D \href{https://doi.org/10.48550/arXiv.1901.07293}{{\bf 99}, 086009} (2019)
	[arXiv:1901.07293 [hep-th]]
	
\bibitem{QT}
	R.~Alicki and R.~Kosloff,
	``Introduction to quantum thermodynamics: History and prospects'',
	in {\it Thermodynamics in the Quantum Regime}, eds.\ F.~Binder et al.,
	(Cham: Springer, 2019), \href{https://doi.org/10.1007/978-3-319-99046-0_1}{pp.\ 1--33}
	[arXiv:1801.08314 [quant-ph]].
	
\bibitem{Sewell}
	G.~L.~Sewell,
	``On the question of temperature transformations under Lorentz and Galilei boosts'',
	J.\ Phys.\ A: Math.\ Theor.\ \href{https://doi.org/10.1088/1751-8113/41/38/382003}{{\bf 41} 382003} (2008)
	[arXiv:0808.0803 [math-ph]]
	
\bibitem{trouble}
	N.~Goheer, M.~Kleban, and L.~Susskind,
	``The Trouble with de Sitter Space'',
	J. High Energy Phys.\ \href{https://doi.org/10.1088/1126-6708/2003/07/056}{{\bf 0307}, 056} (2003)
	[arXiv:hep-th/0212209]
	
%%%%%%%%%%
%%% RELATION TO OTHER APPROACHES
%%%%%%%%%%

\bibitem{Starobinsky}
	A.~A.~Starobinsky,
	``Stochastic de Sitter (inflationary) stage in the early universe'',
	Lect.\ Notes Phys.\ \href{https://doi.org/10.1007/3-540-16452-9_6}{{\bf 246}, 107} (1986)

\bibitem{Mirbabayi}
	M.~Mirbabayi,
	``Markovian dynamics in de Sitter'',
	J.\ Cosmol.\ Astropart.\ Phys.\ \href{https://doi.org/10.1088/1475-7516/2021/09/038}{{\bf 2109}, 038} (2021)
	[arXiv:2010.06604 [hep-th]]

\bibitem{CLPW}
	V.~Chandrasekaran, R.~Longo, G.~Penington, and E.~Witten,
	``An algebra of observables for de Sitter space'',
	 J.\ High Energy Phys.\ \href{https://doi.org/10.1007/JHEP02(2023)082}{{\bf 2302}, 082} (2023)
	[arXiv:2206.10780 [hep-th]

\bibitem{Susskind}
	L.~Susskind,
	``A paradox and its resolution illustrate principles of de Sitter holography'',
	arXiv:2304.00589 [hep-th]
	
\bibitem{Burgess-etal}
	C.~P.~Burgess, R.~Holman, G.~Tasinato, and M.~Williams,
	``EFT beyond the horizon: Stochastic inflation and how primordial quantum fluctuations go classical'',
	 J.\ High Energy Phys.\ \href{https://doi.org/10.1007/JHEP03(2015)090}{{\bf 1503}, 090} (2015)
	[arXiv:1408.5002 [hep-th]]
	
\bibitem{Boyanovsky}
	D.~Boyanovsky,
	``Effective field theory during inflation: Reduced density matrix and its quantum master equation'',
	Phys.\ Rev.\ D \href{https://doi.org/10.1103/PhysRevD.92.023527}{{\bf 92}, 023527} (2015)
	[arXiv:1506.07395 [astro-ph.CO]]
	
\bibitem{Hollowood-McDonald}
	T.~J.~Hollowood and J.~I.~McDonald,
	``Decoherence, discord, and the quantum master equation for cosmological perturbations'',
	Phys.\ Rev.\ D \href{https://doi.org/10.1103/PhysRevD.95.103521}{{\bf 95}, 103521} (2017)
	[arXiv:1701.02235 [gr-qc]]
	
\bibitem{Martin-Vennin}
	J.~Martin and V.~Vennin,
	``Observational constraints on quantum decoherence during inflation'',
	J.\ Cosmol.\ Astropart.\ Phys.\ \href{https://doi.org/10.1088/1475-7516/2018/05/063}{{\bf 1805}, 063} (2018)
	[arXiv:1801.09949 [astro-ph.CO]]
	
\bibitem{Brahma-etal}
	S.~Brahma, A.~Berera, and J.~Calder\'on-Figueroa,
	``Universal signature of quantum entanglement across cosmological distances'',
	Class.\ Quant.\ Grav.\ \href{https://doi.org/10.1088/1361-6382/aca066}{{\bf 39}, 245002} (2022)
	[arXiv:2107.06910 [hep-th]]
	
\bibitem{Colas-etal}
	T.~Colas, J.~Grain, and V.~Vennin,
	``Benchmarking the cosmological master equations'',
	Eur.\ Phys.\ J.\ C \href{https://doi.org/10.1140/epjc/s10052-022-11047-9}{{\bf 82}, 1085} (2022)
	[arXiv:2209.01929 [hep-th]]

%%%%%%%%%%
%%% LOCAL PHYSICS IN EXPANDING SPACE
%%%%%%%%%%
	
\bibitem{Jacobson}
	T.~Jacobson,
	``Introduction to quantum fields in curved spacetime and the Hawking effect'',
	arXiv:gr-qc/0308048
	
\bibitem{Parker-Toms}
	L.~Parker and D.~Toms,
	{\it Quantum Field Theory in Curved Spacetime}
	(Cambridge: Cambridge University Press, 2009)
	
\bibitem{rotating}
	R.~Alicki and A.~Jenkins,
	``Interaction of a quantum field with a rotating heat bath'',
	Ann.\ Phys.\ (NY) \href{https://doi.org/10.1016/j.aop.2018.05.001}{{\bf 395}, 69} (2018)
	[arXiv:1702.06231 [quant-ph]]
	
\bibitem{tribo}
	R.~Alicki and A.~Jenkins,
	``Quantum theory of triboelectricity'',
	Phys.\ Rev.\ Lett.\ \href{https://doi.org/10.1103/PhysRevLett.125.186101}{{\bf 125}, 186101} (2020)
	[arXiv:1904.11997 [cond-mat.mes-hall]]

\bibitem{Price-Romano}
	R.~H.~Price and J.~D.~Romano,
	``In an expanding universe, what doesn't expand?'',
	Am.\ J.\ Phys.\ \href{https://doi.org/10.1119/1.3699245}{{\bf 80} 376} (2012)
	[arXiv:gr-qc/0508052]
	
\bibitem{Faraoni-Jacques}
	V.~Faraoni and A.~Jacques,
	``Cosmological expansion and local physics'',
	Phys.\ Rev.\ D \href{https://doi.org/10.1103/PhysRevD.76.063510}{{\bf 76}, 063510} (2007)
	[arXiv:0707.1350 [gr-qc]]
	
%%%%%%%%%%
%%% MARKOVIAN MASTER EQUATION
%%%%%%%%%%
		
\bibitem{Alicki-Lendi}
	R.~Alicki and K.~Lendi,
	{\it Quantum dynamical semigroups and applications},
	Lect.\ Notes Phys.\ \href{https://doi.org/10.1007/3-540-70861-8}{{\bf 717}, 1} (2007)
	
%%%%%%%%%%
%%% THERMALIZATION IN EXPANDING UNIVERSE
%%%%%%%%%%
	
\bibitem{Feng-Zhang}
	J.~Feng and J.-J.~Zhang,
	``Quantum Fisher information as a probe for Unruh thermality'',
	Phys.\ Lett.\ B \href{https://doi.org/10.1016/j.physletb.2022.136992}{{\bf 827}, 136992} (2022)
	[arXiv:2111.00277 [hep-th]]

\bibitem{BD}
	T.~S.~Bunch and P.~C.~W.~Davies,
	``Quantum field theory in de Sitter space: Renormalization by point-splitting'',
	Proc.\ R.\ Soc.\ London A \href{https://doi.org/10.1098/rspa.1978.0060}{{\bf 360}, 117} (1978)
	
\bibitem{Kolb-Turner}
	E.~Kolb and M.~Turner,
	{\it The Early Universe}
	(Boca Raton: CRC Press, 2018 [1990])
	
\bibitem{Dicke}
	R.~Dicke,
	``Coherence in spontaneous radiation processes'',
	Phys.\ Rev.\ {\bf 93}, 99 (1954)
	
\bibitem{Gross-Haroche}
	M.~Gross and S.~Haroche,
	``Superradiance: An essay on the theory of collective spontaneous emission'',
	Phys.\ Rep.\ \href{https://doi.org/10.1016/0370-1573(82)90102-8}{{\bf 93}, 301} (1982)
	
%%%%%%%%%%
%%% DISCUSSION AND OUTLOOK
%%%%%%%%%%
	
\bibitem{warm1}
	A.~Berera, I.~G.~Moss, and R.~O.~Ramos,
	``Warm inflation and its microphysical basis'',
	Rep.\ Prog.\ Phys.\ \href{https://doi.org/10.1088/0034-4885/72/2/026901}{{\bf 72}, 026901} (2009)
	[arXiv:0808.1855 [hep-ph]]
	
\bibitem{warm2}
	A.~Berera,
	``The warm inflation story'',
	Universe \href{https://doi.org/10.3390/universe9060272}{{\bf 9}, 272} (2023)
	[arXiv:2305.10879 [hep-ph]]
	
\bibitem{Cohen-Kaplan}
	A.~G.~Cohen and D.~B.~Kaplan,
	``Thermodynamic generation of the baryon asymmetry'',
	Phys.\ Lett B \href{https://doi.org/10.1016/0370-2693(87)91369-4}{{\bf 199}, 251} (1987)
	
%%%%%%%%%%
%%% APPENDIX
%%%%%%%%%%
	
\bibitem{historyGKLS}
	D.~Chru\'sci\'nski and S.~Pascazio,
	``A brief history of the GKLS equation'',
	Open Syst.\ Inf.\ Dyn.\ \href{https://doi.org/10.1142/S1230161217400017}{{\bf 24}, 1740001} (2017)
	[arXiv:1710.05993 [quant-ph]]
	
\bibitem{golden}
	R.~Alicki,
	``The Markov master equation and the Fermi golden rule'',
	Int.\ J.\ Theor.\ Phys.\ \href{https://doi.org/10.1007/BF01807150}{{\bf 16}, 351} (1977)
	
\end{thebibliography}
\end{document}